\documentclass[aps,pra,reprint,amsmath,amsfonts,amssymb,superscriptaddress]{revtex4-1}
\usepackage{graphicx,calc}
\usepackage{color,bm}
\usepackage{ulem}
\usepackage{braket}
\usepackage[colorlinks,urlcolor=blue,citecolor=blue,linkcolor=blue]{hyperref}

\begin{document}
	
\title{Machine learning topological invariants of non-Hermitian systems}

\author{Ling-Feng Zhang}
\affiliation{Guangdong Provincial Key Laboratory of Quantum Engineering and Quantum Materials, School of Physics and Telecommunication Engineering, South China Normal University, Guangzhou 510006, China}

\author{Ling-Zhi Tang}
\affiliation{Guangdong Provincial Key Laboratory of Quantum Engineering and Quantum Materials, School of Physics and Telecommunication Engineering, South China Normal University, Guangzhou 510006, China}

\author{Zhi-Hao Huang}
\affiliation{Guangdong Provincial Key Laboratory of Quantum Engineering and Quantum Materials, School of Physics and Telecommunication Engineering, South China Normal University, Guangzhou 510006, China}

\author{Guo-Qing Zhang}\email{zhangptnoone@m.scnu.edu.cn}
\affiliation{Guangdong Provincial Key Laboratory of Quantum Engineering and Quantum Materials, School of Physics and Telecommunication Engineering, South China Normal University, Guangzhou 510006, China}
\affiliation{Guangdong-Hong Kong Joint Laboratory of Quantum Matter, Frontier Research Institute for Physics, South China Normal University, Guangzhou 510006, China}

\author{Wei Huang}
\affiliation{Guangdong Provincial Key Laboratory of Quantum Engineering and Quantum Materials, School of Physics and Telecommunication Engineering, South China Normal University, Guangzhou 510006, China}

\author{Dan-Wei Zhang}\email{danweizhang@m.scnu.edu.cn}
\affiliation{Guangdong Provincial Key Laboratory of Quantum Engineering and Quantum Materials, School of Physics and Telecommunication Engineering, South China Normal University, Guangzhou 510006, China}
\affiliation{Guangdong-Hong Kong Joint Laboratory of Quantum Matter, Frontier Research Institute for Physics, South China Normal University, Guangzhou 510006, China}

\date{\today}

\begin{abstract}
The study of topological properties by machine learning approaches has attracted considerable interest recently. Here we propose  machine learning the topological invariants that are unique in non-Hermitian systems. Specifically, we train neural networks to predict the winding of eigenvalues of {\color{black}four prototypical} non-Hermitian Hamiltonians on the complex energy plane with nearly $100\%$ accuracy. Our demonstrations in the non-Hermitian Hatano-Nelson model, Su-Schrieffer-Heeger model and generalized Aubry-Andr{\'e}-Harper model in one dimension, {\color{black}and two-dimensional Dirac fermion model with non-Hermitian terms} show the capability of the neural networks in exploring topological invariants and the associated topological phase transitions and topological phase diagrams in non-Hermitian systems. Moreover, the neural networks trained by a small data set in the phase diagram can successfully predict topological invariants in untouched phase regions. Thus, our work paves the way to revealing non-Hermitian topology with the machine learning toolbox.
\end{abstract}

\maketitle

\section{Introduction}

Machine learning, which lies at the core of the artificial intelligence and data science, has recently achieved huge success from industrial applications (especially in computer vision and the natural language process) to fundamental researches in physics, cheminformatics and biology~\cite{Jordan2015,lecun2015deep,goodfellow2016deep,carleo2019machine}. In physics, machine learning has shown its availability in experimental data analysis~\cite{Biswas2013,rem2019identifying,Kasieczka2019} and classification of phases of matter~\cite{wang2016discovering,carrasquilla2017machine,zhang2017quantum,PhysRevB.96.195145,huembeli2019automated,dong2019machine,van2017learning,carvalho2018real,zhang2018machine,sun2018deep,huembeli2018identifying,Tsai2019,ming2019quantum,rodriguez2019identifying,Holanda2019,ohtsuki2020drawing}.
Among these applications, one of the most interesting problems is to extract the global properties of topological phases of matter from local inputs, such as the topological invariants that intrinsically nonlocal. Recent works have shown that artificial neural networks can be trained to predict the topological invariants of band insulators with a high accuracy~\cite{zhang2018machine,sun2018deep}. The advantage of this approach is that the neural network can capture global topology directly from local raw data inputs. Other theoretical proposals for identifying topological phases by using supervised or unsupervised learning have been  suggested~\cite{carvalho2018real,huembeli2018identifying,rodriguez2019identifying,Holanda2019,ohtsuki2020drawing,Zhang2019,Long2020,Scheurer2020,PhysRevResearch.2.013354,oleks2020unsupervised}. Notably, the convolutional neural network (CNN) trained from raw experimental data has been demonstrated to identify topological phases \cite{rem2019identifying,lian2019machine}.

On the other hand, growing efforts have been invested in uncovering exotic topological states and phenomena in non-Hermitian systems in recent years~\cite{diehl2008quantum,malzard2015topologically,Lee2016,SYao2018a,SYao2018b,FSong2019,Kunst2018,takata2018photonic,Wang2018,zeng2017anderson,lang2018effects,hamazaki2019non,jin2019bulk,kawabata2019topological,liu2019second,lee2019hybrid,yamamoto2019theory,Hatano1996,Hatano1997,gong2018topological,ghatak2019new,leykam2017edge,shen2018topological,DWZhang2018,GQZhang2020,zhang2020non,Luo2019,jiang2019interplay,longhi2019topological,TLiu2020,HWu2020,QBZeng2020,QBZeng2020b,LZTang2020a,HLiu2020,DWZhang2020b,Xu2020,Liu2020,Xi2019,Lee2020,Yoshida2019}. The non-Hermiticity may come from gain and loss effects~\cite{zeng2017anderson,lang2018effects,takata2018photonic,Wang2018,hamazaki2019non}, non-reciprocal hoppings~\cite{Hatano1996,Hatano1997}, or dissipations in open systems~\cite{diehl2008quantum,malzard2015topologically}. Non-Hermiticity-induced topological phases are also investigated in disordered \cite{zhang2020non,Luo2019,jiang2019interplay,longhi2019topological,TLiu2020,HWu2020,QBZeng2020,QBZeng2020b,LZTang2020a,HLiu2020} and interacting systems~\cite{DWZhang2020b,Xu2020,Liu2020,Xi2019,Lee2020,Yoshida2019}. In non-Hermitian topological systems, there are not only topological properties defined by the eigenstates (such as topological Bloch bands), but also topological invariants solely lying on the eigenenergies. For instance, complex energy landscapes and exceptional points give rise to different topological invariants, which include the winding number (vorticity) defined solely in the complex energy plane \cite{gong2018topological,ghatak2019new,leykam2017edge,shen2018topological}. This winding number and several closely related winding numbers in the presence of symmetries can lead to a richer topological classification than that of their Hermitian counterparts. In addition, it was revealed \cite{PhysRevLett.124.056802,PhysRevLett.124.086801,PhysRevLett.125.126402} that the nonzero winding number in the complex energy plane is the topological origin of the so-called non-Hermitian skin effect \cite{Lee2016,SYao2018a,SYao2018b,FSong2019,Kunst2018}. Considering that the topological invariants in Hermitian systems have been studied recently based on the machine learning approach ~\cite{carvalho2018real,zhang2018machine,sun2018deep,huembeli2018identifying,rodriguez2019identifying,Holanda2019,ohtsuki2020drawing,Zhang2019,Long2020,Scheurer2020}, the flexibility of machine learning such a different kind of winding number in non-Hermitian systems is urgent and meaningful research.

In this work, we adapt machine learning with neural networks to predict non-Hermitian topological invariants and classify the topological phases in several prototypical non-Hermitian models in one and two dimensions. We first take the Hatano-Nelson model \cite{Hatano1996,Hatano1997} as a feasibility verification of machine learning in identifying non-Hermitian topological phases. We show that the trained CNN can predict the winding numbers of eigenenergies with a high accuracy even for those phases that are not included in the training, whereas the fully connected neural network (FCNN) can only predict those in the trained phases. We interpolate the intermediate value of the CNN and find a strong relationship with the winding angle of the eigenenergies in the complex plane. We then use the CNN to study topological phase transitions in a non-Hermitian Su-Schrieffer-Heeger (SSH) model \cite{WPSu1979} with non-reciprocal hopping. We find that the CNN can precisely detect the transition points near the phase boundaries even though trained only by the data in the deep phase region. By using the CNN, we further obtain the topological phase diagram of a non-Hermitian generalized Aubry-Andr\'{e}-Harper (AAH) model \cite{Harper1955,Aubry1980,FLiu2015} with non-reciprocal hopping and a complex quasiperiodic potential. The winding numbers evaluated from the CNN show an accuracy of more than 99\% with theoretical values in the whole parameter space, even though the complex on-site potential is absent in the training process. {\color{black}Finally, we extend our scenario to a two-dimensional non-Hermitian Dirac fermion model \cite{shen2018topological} and show the feasibility of neural networks in revealing the winding numbers associated with exceptional points.} Our work may provide an efficient and general approach to reveal non-Hermitian topology based on the machine learning toolbox.

The rest of this paper is organized as follows. We first study the winding number of the Hatano-Nelson model as a feasibility verification of our machine learning method in Sec.~\ref{sec2}. Different performances of the CNN and the FCNN are also discussed. Section~\ref{sec3} is devoted to revealing the topological phase transition in the non-Hermitian SSH model by the CNN. In Sec.~\ref{sec4}, we show that the CNN can precisely predict the topological phase diagram of the non-Hermitian generalized AAH model. {\color{black}In Sec.~\ref{sec5}, we extend our scenario to reveal the winding numbers associated with exceptional points in a two-dimensional non-Hermitian Dirac fermion model.} A further discussion and short summary are finally presented in Sec.~\ref{sec6}.

\section{\label{sec2}Learning topological invariants in Hatano-Nelson model}

\begin{figure}[!t]
    \includegraphics[width=0.4\textwidth]{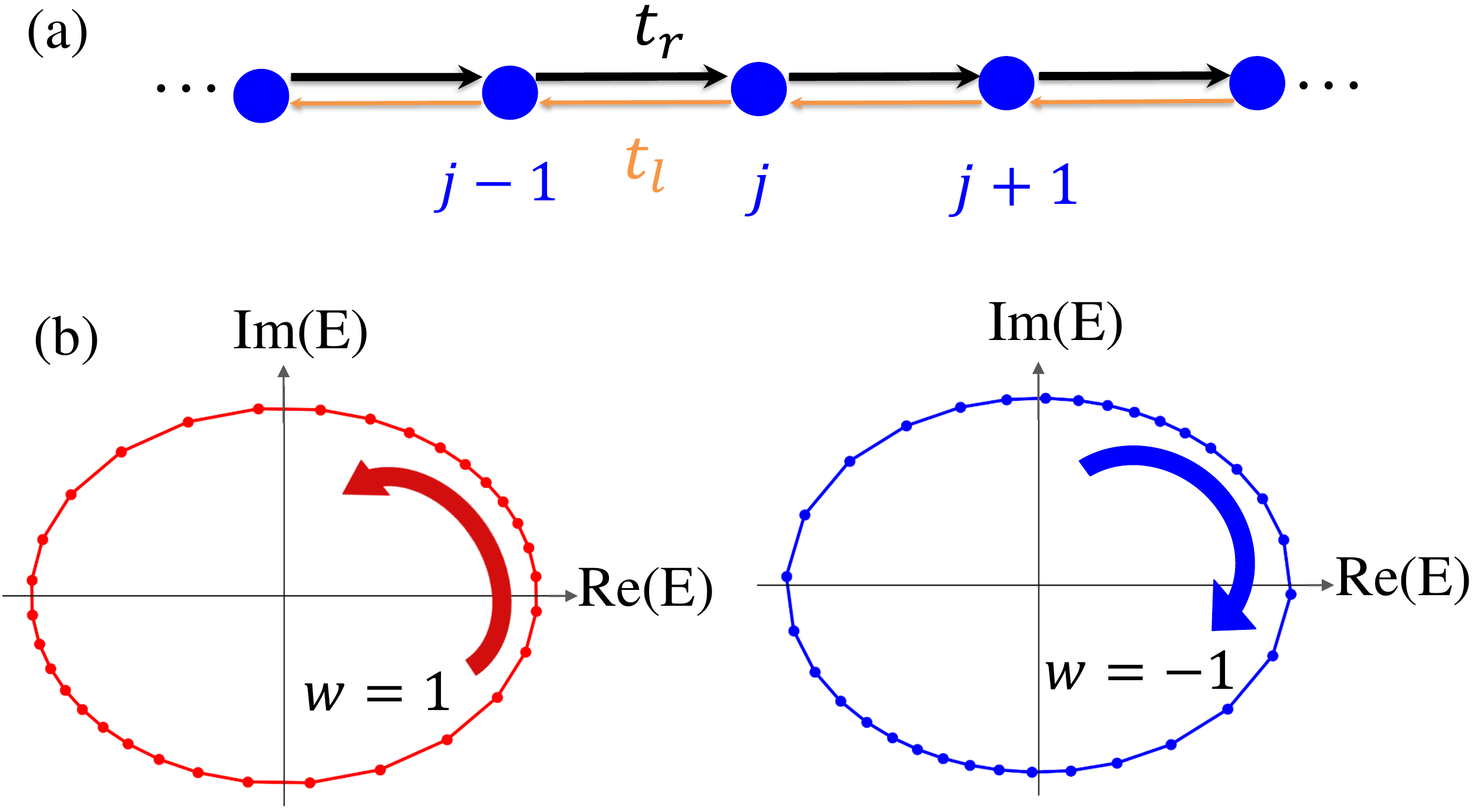}
    \caption{(Color online) (a) The Hatano-Nelson model with non-reciprocal hopping between two nearest-neighbor sites ($\mu=1$). (b) The complex eigenenergy draws a closed loop around the base energy $E_B = 0$ during the variation of quasimomentum $k$ from $0$ to $2\pi$, giving rise to the winding number $w=\pm 1$ for the counterclockwise and clockwise windings, respectively.}
    \label{fig1}
\end{figure}

Let us begin with the Hatano-Nelson model, which is a prototypical single-band non-Hermitian model and takes the following Hamiltonian in a one-dimensional lattice of length $L$ \cite{Hatano1996,Hatano1997}:
\begin{equation}\label{HN-model}
    H_1= \sum_{j}^{L}(t_r \hat{c}^{\dagger}_{j+\mu}\hat{c}_{j}+ t_l \hat{c}^{\dagger}_{j} \hat{c}_{j+\mu} + V_j \hat{c}^{\dagger}_{j} \hat{c}_{j}).
\end{equation}
Here $t_l\neq t_r^*$ denotes the amplitudes of non-reciprocal hopping, $\hat{c}^{\dagger}_{j}(\hat{c}_{j})$ is the creation (annihilation) operator at the $j$-th lattice site, $\mu$ denotes the hopping length between two sites, and $V_j$ is the on-site energy in the lattice. The original Hatano-Nelson model takes the disorder potential with random $V_j$ and the nearest-neighbor hopping with $\mu=1$, as shown in Fig. \ref{fig1}(a). Here we consider the clean case by setting $V_j=0$ and take $\mu$ as a parameter in learning the topological phase transition with neural networks. Under the periodic boundary condition, the corresponding eigenenergies in this case are given by
\begin{equation}
    E_1(k) = \mathcal{H}_1(k) = t_r e^{-i\mu k} + t_l e^{i\mu k},
\end{equation}
where $\mathcal{H}_1(k)$ is the Hamiltonian in momentum space with the quasimomentum $k=0,2\pi/L,4\pi/L,\cdots, 2\pi$.

Following Ref. \cite{gong2018topological}, we can define the winding number in the complex energy place as a topological invariant in the Hatano-Nelson model,
\begin{align}
    w &= \int_{0}^{2\pi}\frac{dk}{2\pi i}\partial_k \ln \det \mathcal{H}_1(k) \nonumber \\
    &=\int_{0}^{2\pi}\frac{dk}{2\pi} \partial_k \arg E_1(k)=\left\{
                                                             \begin{array}{ll}
                                                               \mu, & |t_r|<|t_l|, \\
                                                               -\mu, & |t_r|>|t_l|,
                                                             \end{array}
                                                           \right.
\end{align}
where $\arg$ denotes the principal value of the argument belonging to $[0,2\pi)$. For a discretized $E_1(k)$ with finite lattice site $L$, the complex-energy winding number reduces to
\begin{equation}{\label{eq4}}
    w=\frac{1}{2\pi}\sum_{n=1}^{L}\Delta \theta(n)=\frac{1}{2\pi}\sum_{n=1}^{L}[\theta(n)-\theta(n-1)],
\end{equation}
where $\theta(n)=\arg E_1(2\pi n / L)$. Note that for Hermitian systems ($t_r=t_l^{*}$), one has $w=0$ due to the real energy spectrum with $\arg E_1(k) = 0,\pi$. According to this definition, a nontrivial winding number gives the number of times the complex eigenenergy encircles the base point $E_B=0$, which is unique to non-Hermitian systems. The complex eigenenergy windings for two typical cases with $w=\pm 1$ are shown in Fig. \ref{fig1}(b). To examine whether the neural networks have the ability to learn the winding number in a general formalism, we enable the parameter $\mu$ to control the number of times the complex eigenenergy encircles the origin of the complex plane. When the loop winds around the origin $\mu$ times during the variation of $k$ from $0$ to $2\pi$, the winding number is $\pm \mu$, where $\pm$ indicates counterclockwise and clockwise windings, respectively.

\begin{figure*}[!t]
  \includegraphics[width=0.75\textwidth]{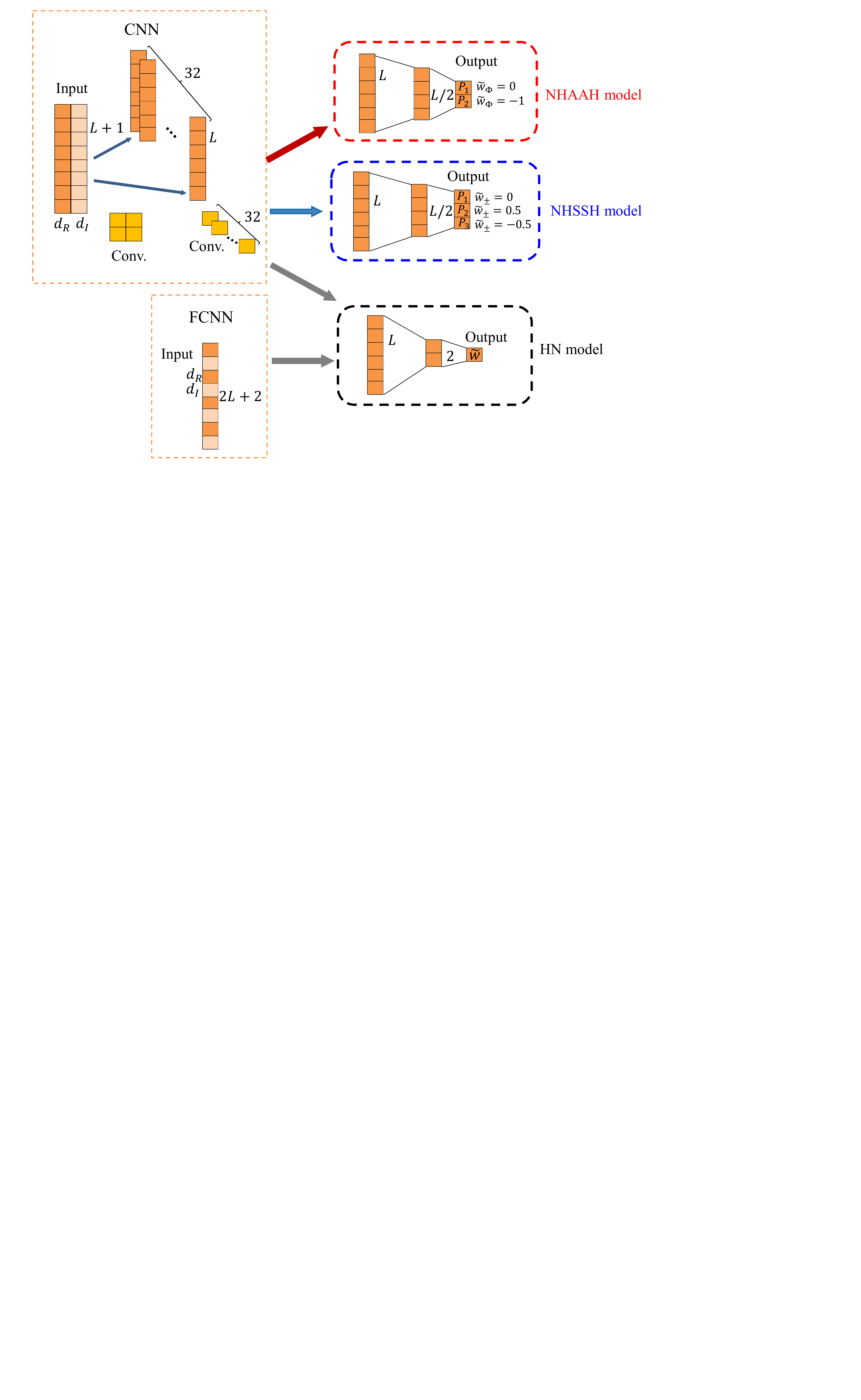}
  \caption{(Color online) Schematic of machine learning workflow and the structure of neural networks for the Hatano-Nelson (HN) model, non-Hermitian SSH (NHSSH) model, and non-Hermitian generalized AAH (NHGAAH) model. The input data are represented by an $(L+1)\times2$-dimensional matrix for the CNN and a $2\times(L+1)$-dimensional vector for the FCNN, respectively. Here $d_R$ and $d_I$ denote the real and imaginary parts of the input data (complex eigenenergies), respectively.}
   \label{fig2}
\end{figure*}

We now build a supervised task for learning the winding number given by Eq. (\ref{eq4}) based on neural networks. First, we need labeled data sets for the training and evaluation. Since the winding number is intrinsically nonlocal and characterized by a complex energy spectrum, we feed neural networks with the normalized spectrum-dependent configurations $\mathbf{d}(n)=[\mathbf{d}_R(n),\mathbf{d}_I(n)]$ at $L$ points discretized uniformly from $0$ to $2\pi$, where $\mathbf{d}_R(n)=\mathrm{Re}[E_1(2\pi n / L)]$ and $\mathbf{d}_I(n)=\mathrm{Im}[E_1(2\pi n / L)]$. Therefore, the input data are an $(L+1)\times2$-dimensional matrix of the form
$$
    \left[ \begin{array}{ccccc}
\mathbf{d}_R(0) & \mathbf{d}_R(2\pi/L) & \mathbf{d}_R(4\pi/L)&... & \mathbf{d}_R(2\pi)\\
\mathbf{d}_I(0) & \mathbf{d}_I(2\pi/L) & \mathbf{d}_I(4\pi/L)&... & \mathbf{d}_I(2\pi)
\end{array}
\right ]^{T},
$$
with a period of $2\pi$: $\mathbf{d}(n) = \mathbf{d}(n+2\pi)$. In the following, we set $L=32$, which is large enough to take discrete energy spectra as the input data of neural networks. Labels are computed according to Eq. (\ref{eq4}) for the corresponding configurations.

The machine learning workflow is schematically shown in Fig. \ref{fig2}. For the Hatano-Nelson model with different $\mu$, the output of the neural network is a real number $\tilde{w}$, and the predicted winding number is interpreted as the integer that is closest to $\tilde{w}$. We first train the neural networks with both complex spectrum configurations and their corresponding true winding numbers. After the training, we feed only the complex-spectrum-dependent configurations to the neural networks and compare their predictions with the true winding numbers, from which we determine the percentage of the correct predictions as the accuracy. In this case, we consider two typical classes of neural networks: the CNN and FCNN, respectively. The neural networks are similar to those in Ref. \cite{zhang2018machine} for calculating the winding number of the Bloch vectors in Hermitian topological bands.

The CNN in our training has two convolution layers with 32 kernels of size $1\times2\times2$ and 1 kernel of size $32\times1\times1$, followed by a fully connected layer of two neurons before the output layer. The total number of trainable parameters is 262. The FCNN has two hidden layers with 32 and 2 neurons, respectively. The total number of trainable parameters is 2213. The architecture of two classes of neural networks is shown in Fig. \ref{fig2}. All hidden layers have rectified linear units $f(x)=\max{(0,x)}$ as activation functions and the output layer has linear activation function $f(x)=x$. The objective function to be optimized is defined by
\begin{equation}
    J_{1}= \frac{1}{N}\sum_{i=1}^{N}(\tilde{w}_i - w_i)^2,
\end{equation}
where $\tilde{w}_i$ and $w_i$ are, respectively, the winding number of the $i$th complex eigenenergies predicted by the neural networks and the true winding number, and $N$ is the total number of the training data set. We take $6\times 10^4$ training configurations, which consist of a ratio of $1:1:1$ of them having winding numbers
$\{\pm1,\pm2,\pm3\}$, respectively. The test set consists of some configurations with winding numbers $w \in \{\pm1,\pm2,\pm3\}$ that are not included in the training set and $w \in \{\pm4,\pm5\}$ that are not seen by neural networks during the training. The number of configurations for each kind of winding number is $4\times10^3$. The training details are given in the Appendix \ref{app}.

\begin{figure}[!t]
    \centering
    \includegraphics[width=0.4\textwidth]{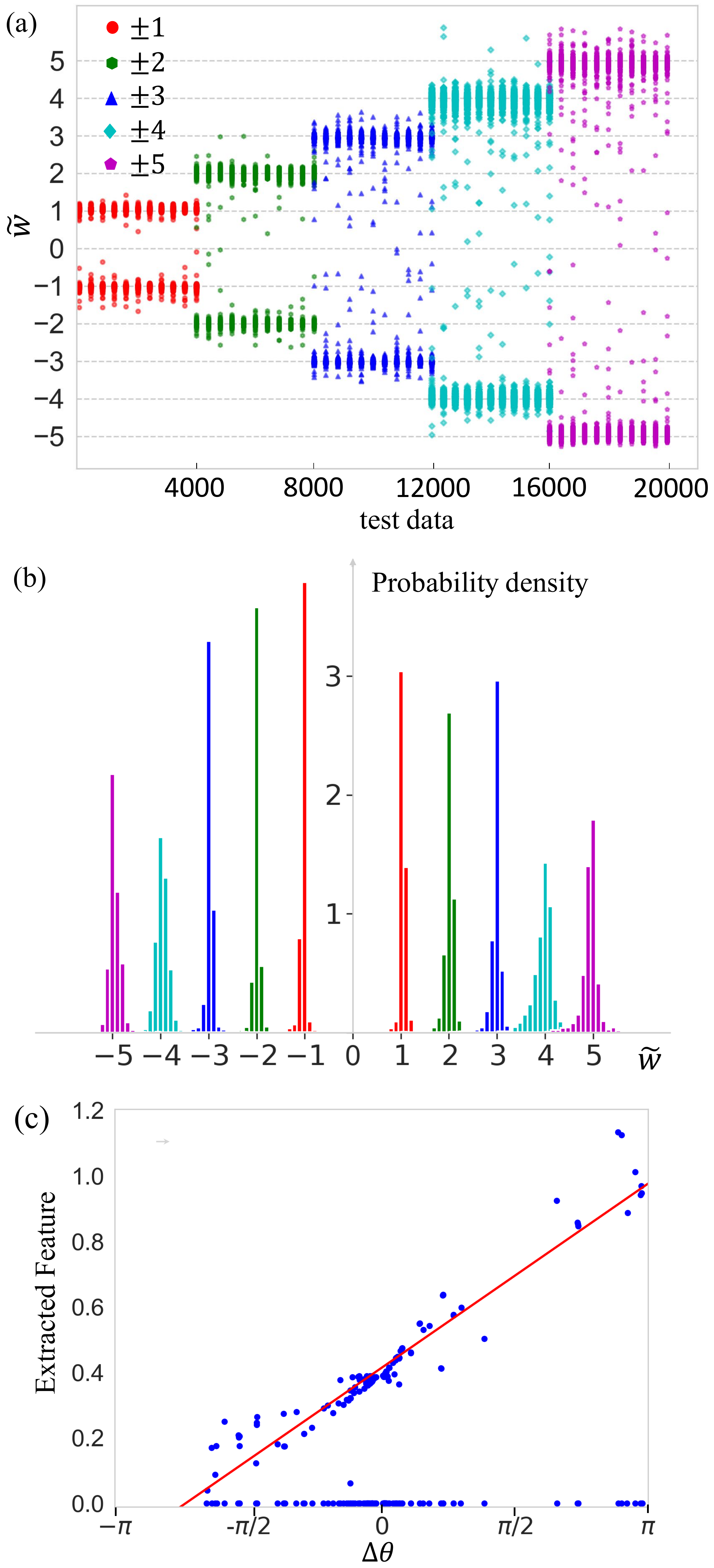}
    \caption{(Color online) (a) Winding numbers predicted by the CNN on test data sets. Different colors represent different true winding numbers, and each test set contains $4\times 10^3$ complex spectrum configurations. (b) Probability distribution of the predicted winding number from the CNN on test data sets. The sum of the probability distribution for a test set (bins of the same color) is equal to 1 and there are narrow peaks at each integer. (c) The intermediate output $a_n$, which is the activation value after two convolutional layers versus the corresponding exact winding angle $\Delta\theta(n)$. $10L$ points corresponding 10 different test configurations are plotted.}
    \label{fig3}
\end{figure}

After training, we test with other configurations and the predicted winding numbers $\tilde{w}$ are shown in Fig. \ref{fig3} (a). Note that the networks tend to produce $\tilde{w}$ close to integers and thus we take each final winding number as the integer closest to $\tilde{w}$. As shown in Fig. \ref{fig3} (b), we plot the probability distribution of $\tilde{w}$ predicted from the CNN on different test data sets. The test results of two neural networks are presented in Table. \ref{tab1}, which shows a very high accuracy (more than $98\%$) of the CNN and FCNN on the test data set with the winding numbers $w=\{\pm1, \pm2, \pm3\}$. We can find that the CNN performs generally better than the FCNN. Surprisingly, the CNN works well even in the cases of $w=\{\pm4,\pm5\}$, which consist of configurations with larger winding numbers not seen by neural networks during the training. On the contrary, the FCNN cannot predict the true winding numbers even though it has more trainable parameters. These results indicate that the convolutional layer respects the translation symmetry of complex spectrum in the momentum space explicitly and convolutional layers can take local winding angle $\Delta \theta$ explicitly through the $2\times 2$ kernels.

To further see the advantage of the CNN, we open up the black box of neural networks and find the relationship between intermediate activation values and physical quantities, i.e. the winding angle $\Delta \theta$. Based on the convolutional layers, we consider that the activation value after two convolutions should have a linear dependence on $\Delta \theta$ to some extent and the following fully-connected layers use a simple linear regression. We plot $a_n$ versus $\Delta\theta(n)$, with $n=1,...,L$ and $a_n$ being the $n$-th component of intermediate values after two convolution layers. As shown in Fig. \ref{fig3} (c), the intermediate output is approximately linear with $\Delta\theta$ within certain regions. A linear combination of these intermediate values with correct coefficients in the following fully connected layers can then easily lead to the true winding number. In this way, the CNN realizes a calculation workflow that is equivalent to the wingding angle $\Delta \theta$ in Eq. (\ref{eq4}).

\begin{table}[tb]
    \centering
    \begin{tabular}{c c c c c c}
\hline\hline
     $w$ & $\pm 1$ & $\pm 2$ & $\pm 3$ & $\mathbf{\pm 4}$ & $\mathbf{\pm 5}$  \\
     \hline
     CNN Accuracy & 99.8 \% & 99.4 \%& 98.0\% & $\mathbf{96.7\%}$ &  $\mathbf{96.0\%}$\\
     \hline
     FCNN Accuracy & 99.2\% & 99.0\% & 98.5\% &$\mathbf{0.0\%}$ &$\mathbf{0.0\%}$\\
     \hline\hline
\end{tabular}{}
    \caption{Accuracy of the CNN and FCNN on test data set with the winding numbers $w= \{\pm1, \pm2, \pm3, \pm4, \pm5\}$ in the Hatano-Nelson model with $\mu=1,2,3,4,5$. The winding numbers $w=\{\pm 4,\pm5\}$ are not seen by the neural networks during the training.} \label{tab1}
\end{table}

\section{\label{sec3}Learning topological transition in non-Hermitian SSH model}

\begin{figure}[!t]
    \centering
    \includegraphics[width=0.4\textwidth]{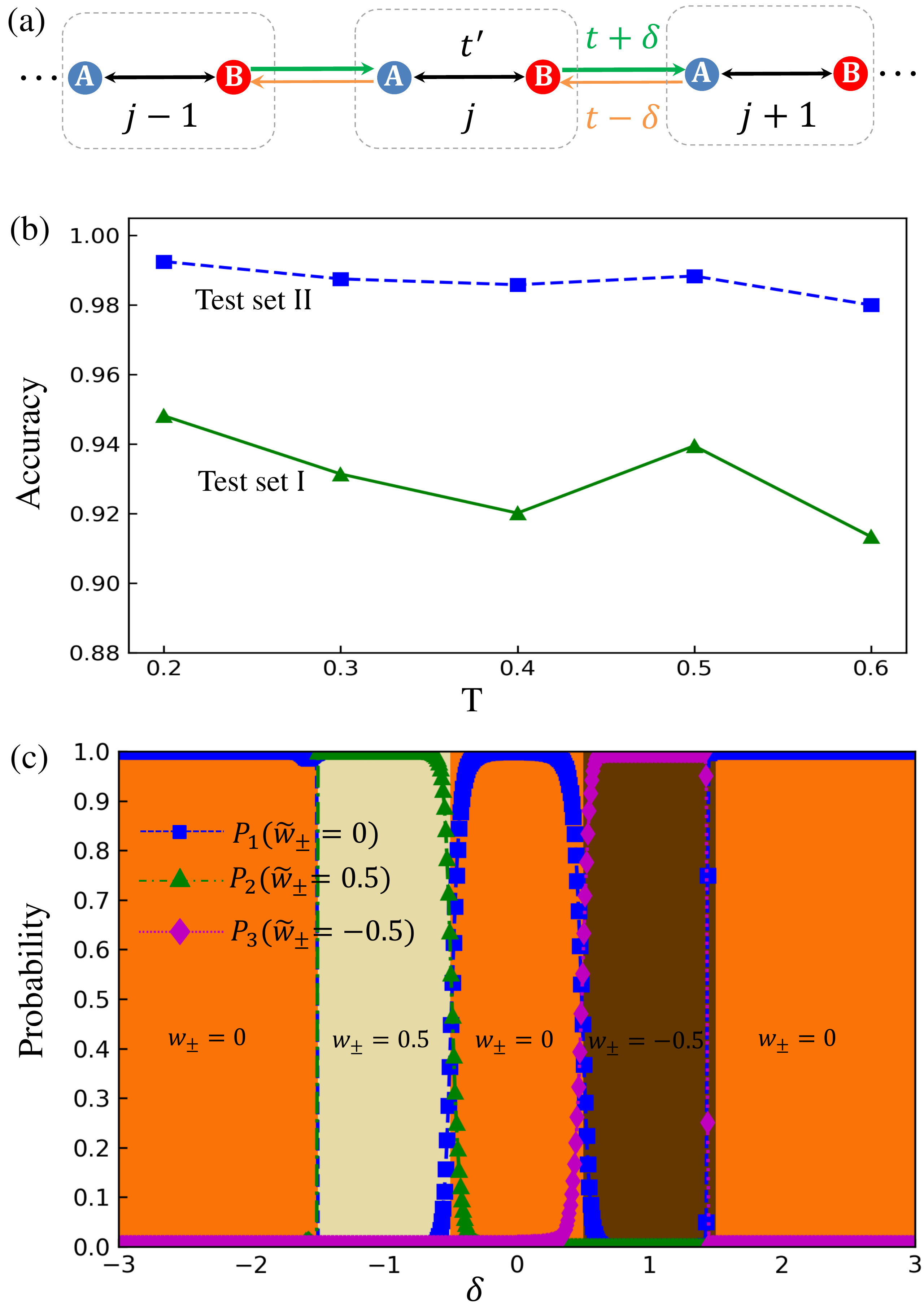}
    \caption{(Color online) (a) The non-Hermitian SSH model with non-reciprocal hopping modulated by the parameter $\delta$. (b) The accuracy of two test sets versus the distance threshold $T$. For each $T$, data sets are regenerated and the CNN is retrained and retested. (c) Classification probabilities outputted by the CNN in test set II with $T=0.2$, where true phase transition points are located at $\delta= \{-1.5,-0.5,0.5,1.5\}$. The predicted phase transition points are located at the crossing point of prediction probabilities. Different colors represent different winding numbers.}
    \label{fig4}
\end{figure}

Based on the accurate winding number calculated by the CNN, we further use a similar CNN to study topological phase transitions in the non-Hermitian SSH model, as shown in Fig. \ref{fig4} (a). The considered model with nonreciprocal intra-cell hopping in the one-dimensional dimerized lattice of $L$ unit cells can be described by the following Hamiltonian:
\begin{equation}
    H_2 = \sum_{n=1}^{L}[(t-\delta)\hat{a}^{\dagger}_{n}\hat{b}_{n}+(t+\delta)\hat{b}^{\dagger}_{n}\hat{a}_{n}+t'\hat{a}^{\dagger}_{n+1}\hat{b}_{n}+t'\hat{b}^{\dagger}_{n}\hat{a}_{n+1}].
\end{equation}
Here $\hat{a}^{\dagger}_{n}$ and $\hat{b}^{\dagger}_{n}$ ($\hat{a}_{n}$, $\hat{b}_{n}$) denote the creation (annihilation) operators on the $n$-th $A$ and $B$ sublattices, $t$ is the uniform intra-cell hopping amplitude, $\delta$ is the non-Hermitian parameter, and $t'$ is the inter-cell hopping amplitude. When $\delta=0$, this model reduces to the Hermitian SSH model. Under the periodic boundary condition, the corresponding Hamiltonian in the momentum space is given by
\begin{equation}
    \mathcal{H}_2(k) = \left( \begin{array}{cc}
0 & t'e^{-ik}+t-\delta\\
t'e^{ik}+t+\delta & 0 \end{array}
\right ).
\end{equation}
The two energy bands are then given by
\begin{equation}
    E_{\pm}(k) = \pm \sqrt{1+t^{2}-\delta^{2}+2t\cos{k}-i2\delta\sin{k}}.
\end{equation}
Following Ref. \cite{leykam2017edge,shen2018topological,gong2018topological,ghatak2019new,PhysRevX.9.041015} and considering the sublattice symmetry, one can define an inter-band winding number
\begin{equation}
    w_{\pm}=\int_{0}^{2\pi}\frac{dk}{2\pi}\partial_k \arg(E_{+} - E_{-}) =\int_{0}^{2\pi}\frac{dk}{4\pi}\partial_k\arg E_{+}^2.
\end{equation}
For discretized $E_{\pm}(k)$ with finite $L$, it reduces to
\begin{equation}
    w_{\pm} = \frac{1}{4\pi} \sum_{n=1}^{L}[\theta'(n)-\theta'(n-1)],
\end{equation}
with $\theta'(n)=\arg{E_{+}^2(2\pi n/L)}$ in this model. Notably, $w_{\pm}$ is half the total windings of $t'e^{-ik}+t-\delta$ and $t'e^{ik}+t+\delta$ around the origin of the complex plane as $k$ is increased from $0$ to $2\pi$. The inter-band winding number $w_{\pm}$ is quantized as $\mathbb{Z}/2$ because the windings of $t'e^{-ik}+t-\delta$ and $t'e^{ik}+t+\delta$ are always integers due to periodicity \cite{shen2018topological}. We consider $t'=1$, $t\in(-6,6)$, and $\delta\in(-6,6)$ in our study.

For this model, we set the configuration of input data as $\mathbf{d}(n)=\{\mathrm{Re}[E_{+}^2(2\pi n / L)], \mathrm{Im}[E_{+}^2(2\pi n / L)]\}$. To learn the topological phase transition in this model, we treat it as a classification task assisted by neural networks. The output of the neural network is the probabilities of different winding numbers. We define $\{P_1, P_2, P_3\}$ as the output probabilities of winding numbers $\tilde{w}_{\pm}=\{0, 0.5, -0.5\}$, respectively. The predicted winding number is interpreted as $\tilde{w}_{\pm}$, which has the highest probability. The architecture of the CNN is shown in Fig. \ref{fig2}, with some training details given in the Appendix \ref{app}. For our task, the objective function to be optimized is defined by
\begin{equation}
    J_{2} = -\frac{1}{N}[\sum_{i=1}^{N}\sum_{j=1}^{n_w=3}1(w_{\pm}^{(i)}=\tilde{w}_{\pm,j})\log_2(P_j))],
    \label{eq12}
\end{equation}
where $w_{\pm}^{(i)}$ is the label of the $i$th configuration, and the set $\{\tilde{w}_{\pm,1}, \tilde{w}_{\pm,2},...,\tilde{w}_{\pm,n_w}\}$ represents the winding number predicted by the neural networks. The expression $1(w_{\pm}^{(i)}=\tilde{w}_{\pm,j})$ means that it will take the value 1 when the condition ${w_{\pm}^{(i)}}=\tilde{w}_{\pm,j}$ is satisfied and the value 0 in the opposite case. In this model, $n_w=3$ and $\{\tilde{w}_{\pm,1}, \tilde{w}_{\pm,2}, \tilde{w}_{\pm,3}\}$ represent the winding numbers $w=\{0, 0.5, -0.5\}$ correspondingly.

To see whether the CNN is a good tool to study topological phase transitions in this model, we define a Euclidean distance $s$ between the configuration and the phase boundaries in the parameter space of the Hamiltonian:
\begin{equation}
    s=\frac{|A\delta+B t+C|}{\sqrt{A^2+B^2}},
\end{equation}
where $A\delta+B t +C=0$ (straight lines in the parameter space about $\delta$ and $t$) is the equation of phase boundaries with $A, B, C$ being the parameters of the equation. In addition, we define a distance threshold $T$. In the following, we choose $T=0.2$ as a demonstration and the situation of $0.2< T\leq 0.6$ is discussed later. The training data set consists of $2.4\times 10^4$ configurations satisfying $s\geq T$ that are sampled from different phases with different winding numbers.

We test the CNN with two test data sets: (I) $6\times 10^3$ configurations satisfying $s< T$ and (II) 300 configurations distributed uniformly in $t=0.5, \delta=[-3,3]$. The data sets distribution and some training details are given in the Appendix \ref{app}. After the training, both test data sets, I and II, are evaluated by the CNN. We use the same training and test workflow for $T =0.3, 0.4,0.5, and 0.6$. Figure \ref{fig4} (b) shows the accuracy of the test data sets versus the distance threshold $T$. We find that the CNN achieves a high accuracy in different $T$, meaning that the CNN can detect the phase transitions precisely in these regions. Moreover, we locate the phase transition points from the crossing points of prediction probabilities; the phase transitions determined by this method are relatively accurate, as shown in Fig. \ref{fig4} (c). In the deep phase, the probability for the true winding number $w_{\pm}$ stays at nearly $100\%$ . On the other hand, the probability for $w_{\pm}$ increases linearly at the phase transitions. In a word, the CNN is a great supplementary tool to study the phase transitions when only phase properties in some confident regions (e.g., the deep phase) are provided.

\begin{figure}[!t]
    \includegraphics[width=0.4\textwidth]{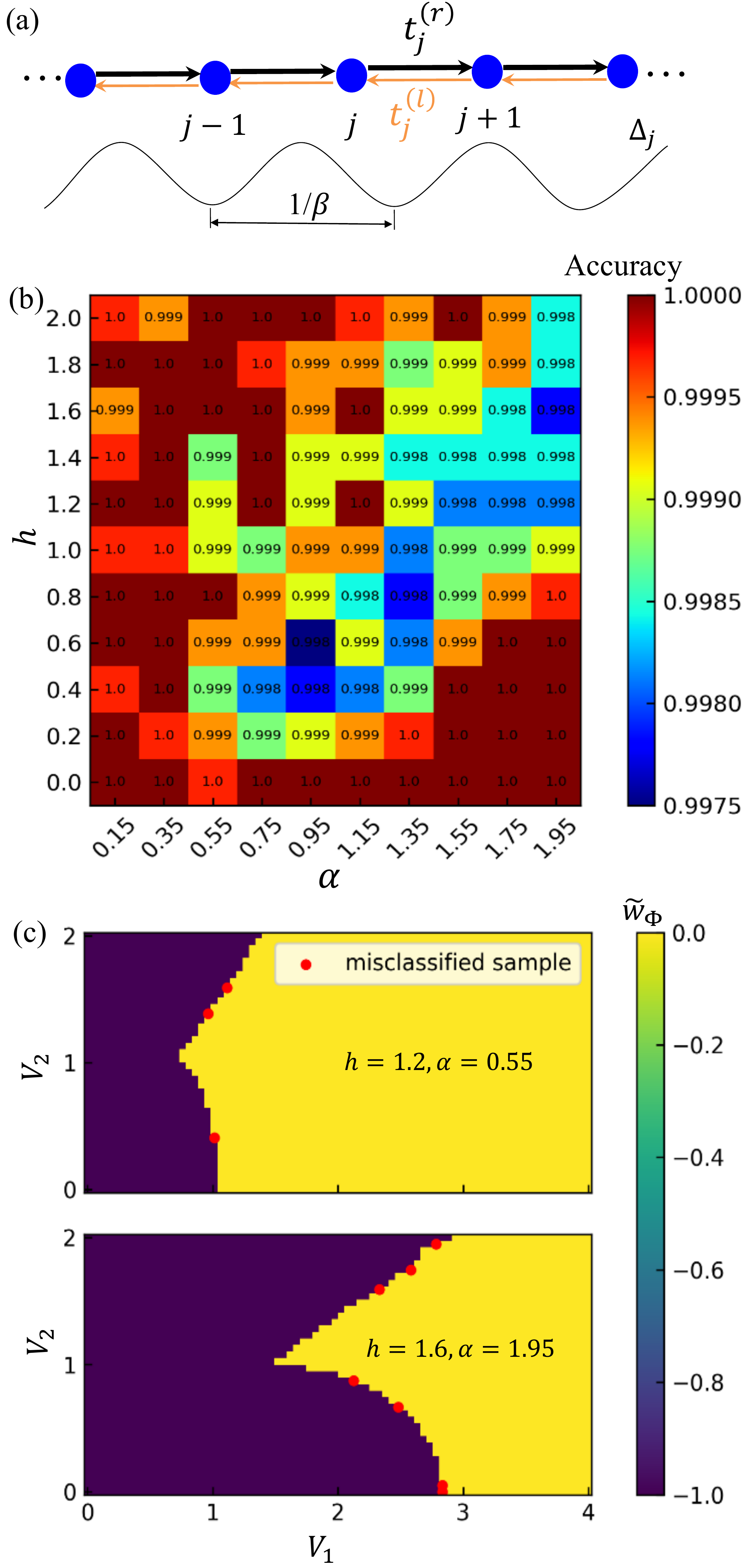}
    \caption{(Color online) (a) Non-Hermitian generalized AAH model with non-reciprocal hopping and a complex quasiperiodic potential. (b) Test accuracy table with respect to two non-Hermiticity parameters, $\alpha$ and $h$. (c) The upper (lower) figure is the topological phase diagram predicted by the CNN for $h=1.2$ and $\alpha=0.55$ ($h=1.6$ and $\alpha=1.95$). Misclassified samples are distributed on the topological transition boundary.}
    \label{fig5}
\end{figure}

\section{\label{sec4}Learning topological phase diagram in non-Hermitian AAH model}

To show that our results can be generalized to other non-Hermitian topological models, we consider a generalized AAH model in a one-dimensional quasicrystal as shown in Fig. \ref{fig5} (a), with two kinds of non-Hermiticities arising from the nonreciprocal hopping \cite{jiang2019interplay} and complex on-site potential phase \cite{longhi2019topological}. The Hamiltonian of such a non-Hermitian AAH model is given by \cite{Tang2020}
 \begin{align}
    H_3=\sum_{j}(t^{(r)}_{j}\hat{c}^{\dagger}_{j+1}\hat{c}_{j}+t^{(l)}_{j}\hat{c}^{\dagger}_{j}\hat{c}_{j+1}+\Delta_j\hat{n}_j),
\end{align}
where the non-reciprocal hopping terms and the on-site potential are parameterized as
\begin{align}
t^{(r)}_{j} &=\{t+V_2\cos[2\pi(j+1/2) \beta]\}e^{-\alpha}, \nonumber \\
t^{(l)}_{j}&=\{t+V_2\cos[2\pi(j+1/2)\beta]\}e^{\alpha}, \\
\Delta_j&=V_1\cos{(2\pi j\beta+ih)}. \nonumber
\end{align}
Here $t^{(r)}_{j}$ $(t^{(l)}_{j})$ denotes the right-hopping (left-hopping) amplitude between the $j$-th and the $(j+1)$-th site with parameters $t>0$ and $V_2$ being real, $\Delta_j$ denotes the complex quasiperiodic potential with $V_1>0$ and $\beta$ an irrational number, and the parameters $\alpha$ and $h$ tune the non-reciprocity and complex phase, respectively. For finite quasiperiodic systems, one can take the lattice site number $L=F_{j+1}$ as a rational number and $\beta=F_{j}/F_{j+1}$ with $F_{j}$ being the $j$-th Fibonacci number since $\lim_{j\rightarrow\infty}F_{j}/F_{j+1} = (\sqrt{5}-1)/2$. In the following, we set $t=1$ and $L=89$.

The winding numbers discussed previously cannot be directly used here due to the periodicity breaking. In this case, one can consider a ring chain with an effective magnetic flux $\Phi$ penetrating through the center, such that the Hamiltonian matrix can be rewritten as
\begin{equation}
    H_3(\Phi)=\left(\begin{array}{ccccc}
    \Delta_1&t^{(l)}_{1}& & &t^{(r)}_{L}e^{-i\Phi}\\t^{(r)}_{1}&\Delta_2&t^{(l)}_{2}& & \\ &\ddots&\ddots&\ddots& \\ & &t^{(r)}_{L-2}&\Delta_{L-1}&t^{(r)}_{L-1}\\t^{(l)}_{L}e^{i\Phi}& & &t^{(r)}_{L-1}&\Delta_L
    \end{array} \right).
\end{equation}
One can define the winding number with respect to $\Phi$ and the energy base $E_B$ \cite{gong2018topological,jiang2019interplay}:
\begin{equation}
    w_{\Phi} = \int_{0}^{2\pi}\frac{\mathrm{d} \Phi}{2\pi i}\partial_{\Phi}\ln\det[{H_3(\Phi)-E_B}].
\end{equation}
 Here $w_\Phi$ counts the number of times the complex spectral trajectory encircles the energy base $E_B$ ($E_B\in \mathbb{C}$ does not belong to the energy spectrum) when the flux $\Phi$ varies from $0$ to $2\pi$. For discretized $H_3(\Phi)$ with $\Phi=0,2\pi/L_{\Phi},4\pi/L_{\Phi},\cdots, 2\pi$, the winding number can be rewritten as
\begin{equation}
    w_{\Phi} = \frac{1}{2\pi} \sum_{n=1}^{L_{\Phi}}[\theta''(n)-\theta''(n-1)],
\end{equation}
where $\theta''(n)=\arg{\det [H_3(2\pi n/L_{\Phi})-E_B]}$.

Below we show that the generalization ability enables the CNN to precisely obtain topological phase diagrams of this non-Hermitian generalized AAH model, even though we only use nonreciprocal-hopping configurations in the training. To do this, we treat the problem as a classification task and set the configuration in this case as $\mathbf{d}(n)=\{\mathrm{Re}\det[\tilde{H}_3(n)], \mathrm{Im}\det[\tilde{H}_3(n)]\}$ with $\tilde{H}_3(n)\equiv H_3(2\pi n/L_{\Phi})-E_B$. The architecture of the CNN is similar to that of the non-Hermitian SSH model, but the output layer now becomes two neurons for two kinds of winding numbers. We define $\{P_1, P_2\}$ as the output probabilities of the winding numbers $\tilde{w}_{\Phi}=\{0, -1\}$, respectively. The objective function in this case is given by [similarly to that in Eq. (\ref{eq12})]
\begin{equation}
    J_{3} = -\frac{1}{N}[\sum_{i=1}^{N}\sum_{j=1}^{n_w=2}1(w_{\Phi}^{(i)}=\tilde{w}_{\Phi,j})\log_2(P_j))],
\end{equation}
where $\{\tilde{w}_{\Phi,1}, \tilde{w}_{\Phi,2}\}$ (with $n_w=2$) represent $\tilde{w}_{\Phi}=\{0, -1\}$, respectively.

To test the generality of the neural network, we train the neural network with configurations corresponding to Hamiltonians with $h=0$, and test it by using configurations corresponding to Hamiltonians with both nonreciprocal hopping amplitudes ($\alpha\neq0$) and complex potentials ($h\neq0$). The training data set includes configurations with $\alpha\in[0.1,1.0]$ and the interval $\Delta \alpha=0.1$; each one consists of $3.2\times10^3$ configurations corresponding Hamiltonians sampled from the two-dimensional parameter space spanned by $V_1\in[0,4]\times V_2\in[0,2]$. The test data set includes 110 pairs of parameters, which consist of $\alpha$ from $\alpha=0.15$ to $\alpha=1.95$ with the interval $\Delta \alpha=0.2$ and $h$ from $
h=0.0$ to $h=2.0$ with the interval $\Delta h=0.2$. We sample $3.2\times10^3$ configurations corresponding to Hamiltonians from the region $V_1\in[0,4]\times V_2\in[0,2]$ for each pair of parameters.

After the training, we find that the CNN performs well even without knowledge of the complex on-site potential ($h=0$) during the training process. Figure \ref{fig5}(b) shows the test accuracy table with respect to the two non-Hermiticity parameters $\alpha$ and $h$, with the accuracy more than $99\%$ in the whole parameter region. Moreover, we present the topological phase diagrams with respect to $V_{1}$ and $V_{2}$ predicted by the CNN, as shown in Fig. \ref{fig5}(c). It is clear that the CNN performs excellently in the deep phase with only a little struggle near the topological phase transitions. We attribute the high accuracy in this learning task to two factors. First, the normalizing data enable both the training and the test data distribution in the complex unit, which is important for the generality of the neural network. Second, the topological transitions in this model are consistent with the real-complex transitions in the energy spectrum \cite{Tang2020}, which reduces the complexity of the problem when input data are dependent on a complex spectrum.

{\color{black} \section{\label{sec5}Generalization to two-dimensional model}

Previously, we have used neural networks to investigate the topological properties of several non-Hermitian models in one dimension. In this section, we extend our scenario to reveal the winding numbers associated with exceptional points in the two-dimensional non-Hermitian Dirac fermion model proposed in Ref. \cite{shen2018topological}. The Dirac Hamiltonian with non-Hermitian terms in two-dimensional momentum space $\mathbf{k}=(k_x,k_y)$ is given by \cite{shen2018topological}
\begin{equation}
    \mathcal{H}_4(\mathbf{k})=(k_x+i\kappa_x)\sigma_x+(k_y+i\kappa_y)\sigma_y+(m+i\delta_m)\sigma_z,
\end{equation}
where $\sigma_{x,y,z}$ are the Pauli matrices, $\kappa_{x,y}$ denote the non-Hermitian modulation parameters, and $m$ and $\delta_m$ denotes the real and imaginary parts of the Dirac mass, respectively.   The corresponding energy dispersion is obtained as \begin{equation}
    E_{\pm}(\mathbf{k})=\pm\sqrt{k^2-\kappa^2+m^2-\delta_m^2+2i(\mathbf{k}\cdot\bm{\kappa})+m\delta_m},
\end{equation}
where $k\equiv|\mathbf{k}|$, $\bm{\kappa}\equiv(\kappa_x,\kappa_y)$ and $\kappa\equiv|\bm{\kappa}|$. The inter-band winding number $w_{\pm}(\Gamma)$ is defined for the energies $E_+(\mathbf{k})$ and $E_-(\mathbf{k})$ in the complex energy plane \cite{shen2018topological}:
\begin{equation}
    w_{\pm}(\Gamma) = \oint_\Gamma \frac{dk}{2\pi}\partial_{\mathbf{k}}\arg{[E_+(\mathbf{k})-E_-(\mathbf{k})]},
\end{equation}
where $\Gamma$ is a closed loop in the two-dimensional momentum space. A nonzero winding number $w_{\pm}(\Gamma)$ implies a band degeneracy in the region enclosed by $\Gamma$. For a pair of separable bands, the winding number can be nonzero only for non-contractible loops in the momentum space. Here we choose loop $\Gamma$ as a unit circle that encircles an exceptional point (a band degeneracy in non-Hermitian band structures) when the Hamiltonian has exceptional points; otherwise we randomly choose a closed loop. The exact topological phase diagram \cite{shen2018topological} in the parameter space spanned by ($m,\kappa$) is shown in Fig.~\ref{fig6}(a). The winding number is 0 in the regime $\kappa<|m|$, and the corresponding Hamiltonian has a pair of separable bands without band degeneracies. In the regime $\kappa>|m|$, the two bands $E_{\pm}(\mathbf{k})$ cross at two isolated exceptional points $\mathbf{k}_{\pm}$ in the two-dimensional momentum space \cite{shen2018topological}
\begin{equation}
    \mathbf{k}_{\pm} = -\frac{m\delta_m}{\kappa}\hat{\mathbf{n}}\pm\frac{\sqrt{(\kappa^2-m^2)(\kappa^2+\delta^2_m)}}{\kappa}\hat{\mathbf{z}}\times\hat{\mathbf{n}},
\end{equation}
where $\hat{\mathbf{n}}=\bm{\kappa}/\kappa$. For the regime $\kappa> |m|$, the inter-band winding numbers $w_{\pm}(\Gamma)$ circling an exceptional point are half-integers and have opposite signs for $\mathbf{k}_{\pm}$. Thus, the winding number $w_{\pm}(\Gamma)$ associated with the exceptional points characterizes topological phase transitions in this model. Note that here we consider the loop $\Gamma$ clockwise circling the exceptional point $\mathbf{k}_{+}$ for the two energy bands $E_{\pm}(\mathbf{k})$ in the complex plane, as displayed in Fig.~\ref{fig6}(b).

\begin{figure}[!t]
    \includegraphics[width=0.48\textwidth]{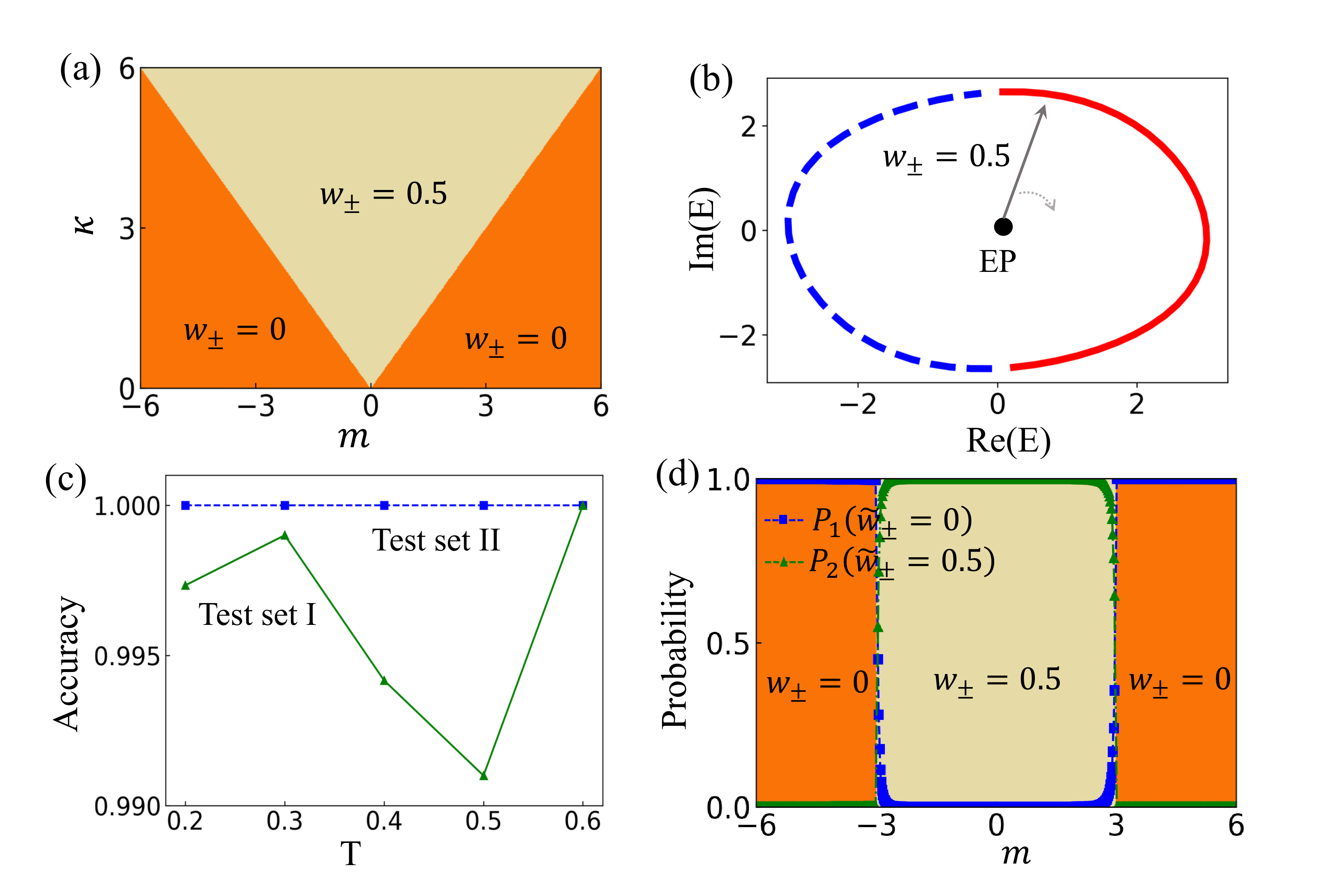}
    \caption{\color{black}(Color online) (a) Phase diagram of the two-dimensional non-Hermitian Dirac fermion model with $\delta_m=1$ and $\bm{\kappa}=(\kappa_x,0)$. (b) The pair switching of
eigenvalues $E_{\pm}(\mathbf{k})$ (denoted by solid red and dashed blue lines) around an exceptional point ($\mathbf{k}_{+}$; denoted by EP) gives rise to the winding number $w_{\pm}(\Gamma)=0.5$. (c) The accuracy of two test sets versus the distance threshold $T$. For each $T$, the data sets are regenerated and the CNN is retrained and retested. (d) Classification probabilities outputted by the CNN in the test set II with $\kappa=3$ and $T=0.5$. The predicted topological phase transition points ($m=\pm3$) located at the crossing points of prediction probabilities. Different colors represent different winding numbers.}
    \label{fig6}
\end{figure}

In the training, we discretize the loop $\Gamma$ to $L$ equally distributed points and set the configuration of input data as $\mathbf{d}(n)=\{\mathrm{Re}{\Delta E(n)},\mathrm{Im}{\Delta E(n)}\}$ with $\Delta E(n)=E_{+}(\mathbf{k}_n)-E_{-}(\mathbf{k}_n)$. The corresponding winding numbers are used as the data labels. We use a workflow similar to that described in Sec.~\ref{sec3} and a CNN with the same structure as described in Sec.~\ref{sec4} to study topological phase transitions characterized by $w_{\pm}$ in this two-dimensional non-Hermitian model. The training data set consists of $3\times10^4$ configurations satisfying $s \geq T$, sampled from different phases with different winding numbers. We test the CNN with two test data sets: (I) $6\times 10^3$ configurations satisfying $s < T$ and (II) 600 configurations distributed uniformly in $\kappa=3, m\in[-6,6]$.
The CNN evaluates both the test data sets, I and II, after the training. In Fig.~\ref{fig6}(c), we plot the accuracy versus the distance threshold $T$, where the CNN is able to detect the winding number precisely for different thresholds $T$. Furthermore, the topological phase transitions can be revealed by the crossing points of the prediction probabilities as shown in
Fig.~\ref{fig6}(d). These results demonstrate the feasibility of neural networks in learning the topological invariants in two-dimensional non-Hermitian models.

}

\section{\label{sec6}Conclusions}

In summary, we have demonstrated that artificial neural networks can be used to predict the topological invariants and the associated topological phase transitions and topological phase diagrams in four different non-Hermitian models with a high accuracy. The eigenenergy winding numbers in the Hatano-Nelson model are presented as a demonstration of our machine learning method. The CNN trained by the data set within the deep phases has been shown to correctly detect the phase transition near each boundary of the non-Hermitian SSH model. We have also investigated the non-Hermitian generalized AAH model with non-reciprocal hopping and a complex quasiperiodic potential. It is found that the topological phase diagram in the non-Hermiticity parameter space predicted by the CNN has a high accuracy with the theoretical counterpart. Furthermore, we have generalized our scenario to reveal the winding numbers associated with exceptional points in the two-dimensional non-Hermitian Dirac fermion model. Our results have shown the generality of the machine learning method in classifying topological phases in prototypical non-Hermitian models.

{\color{black} Finally, we make some remarks on future studies on machine learning non-Hermitian topology. Some exotic features of non-Hermitian topological systems are sensitive to the boundary condition, such as the non-Hermitian skin effect under open boundary conditions \cite{Lee2016,SYao2018a,SYao2018b,FSong2019,Kunst2018}, which is closely related to the winding number of complex eigenenergies \cite{PhysRevLett.125.126402,PhysRevLett.125.226402,PhysRevLett.124.086801}. The energy spectrum under periodic boundary conditions may deviate drastically from that under open boundary conditions. Further studies on the non-Hermitian skin effects and the classification of non-Hermitian topological phases under open boundary conditions based on machine learning algorithms will be conducted. In addition, machine learning non-Hermitian topological invariants defined by the eigenstates would be an interesting further study.

\textit{Note added.} Recently, we noticed two related works on machine learning non-Hermitian topological states~\cite{narayan2020machine,yu2020unsupervised}, which focused on the winding number of the Hamiltonian vectors and the cluster of non-Hermitian topological phases in an unsupervised fashion, respectively.
}

\appendix

\begin{figure}[!t]
    \includegraphics[width=0.4\textwidth]{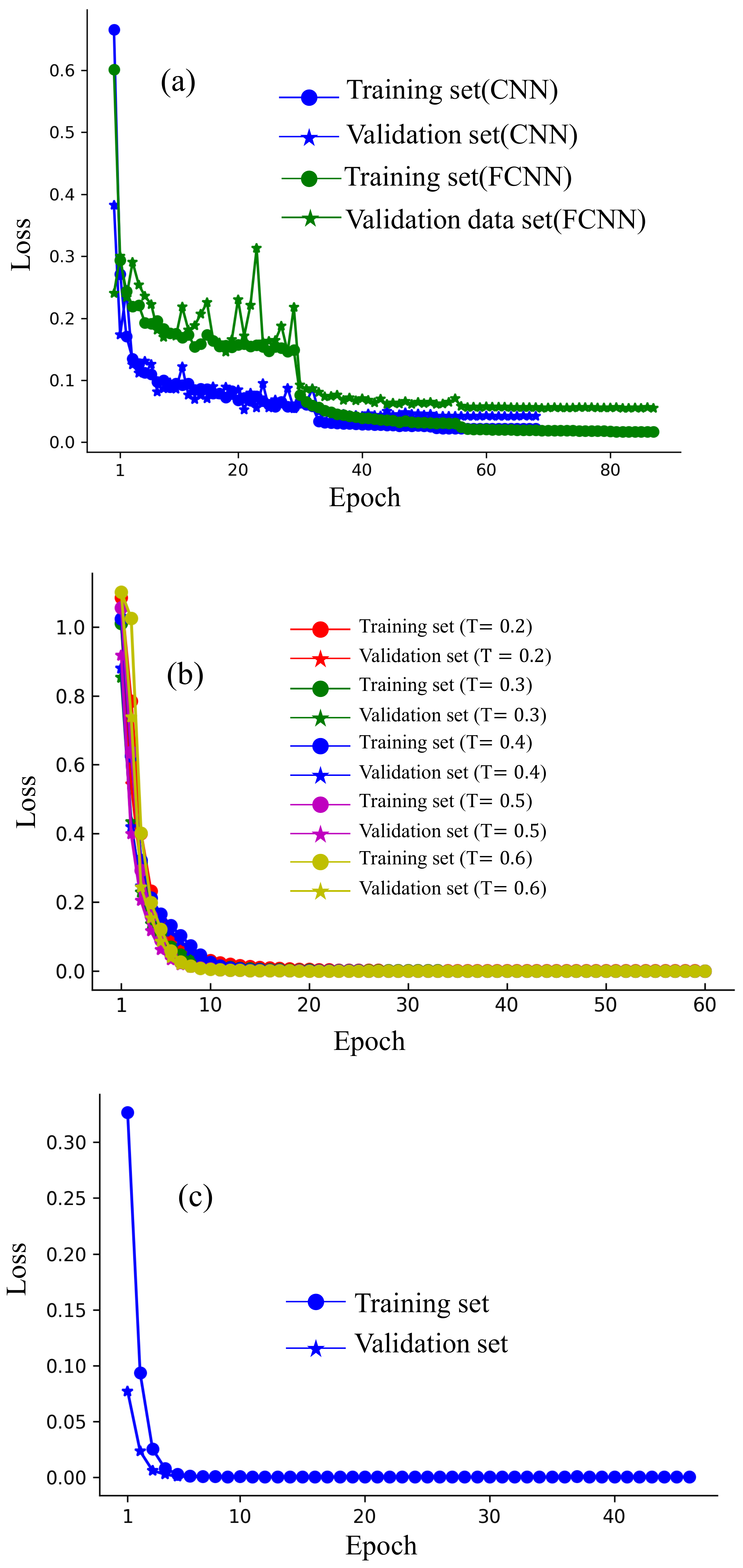}
    \caption{(Color online) (a) CNN and FCNN training loss history in the Hatano-Nelson model. CNN training loss history in (b) the non-Hermitian SSH model; and (c) the non-Hermitian generalized AAH model.}
    \label{fig7}
\end{figure}

\begin{figure}[!t]
    \includegraphics[width=0.4\textwidth]{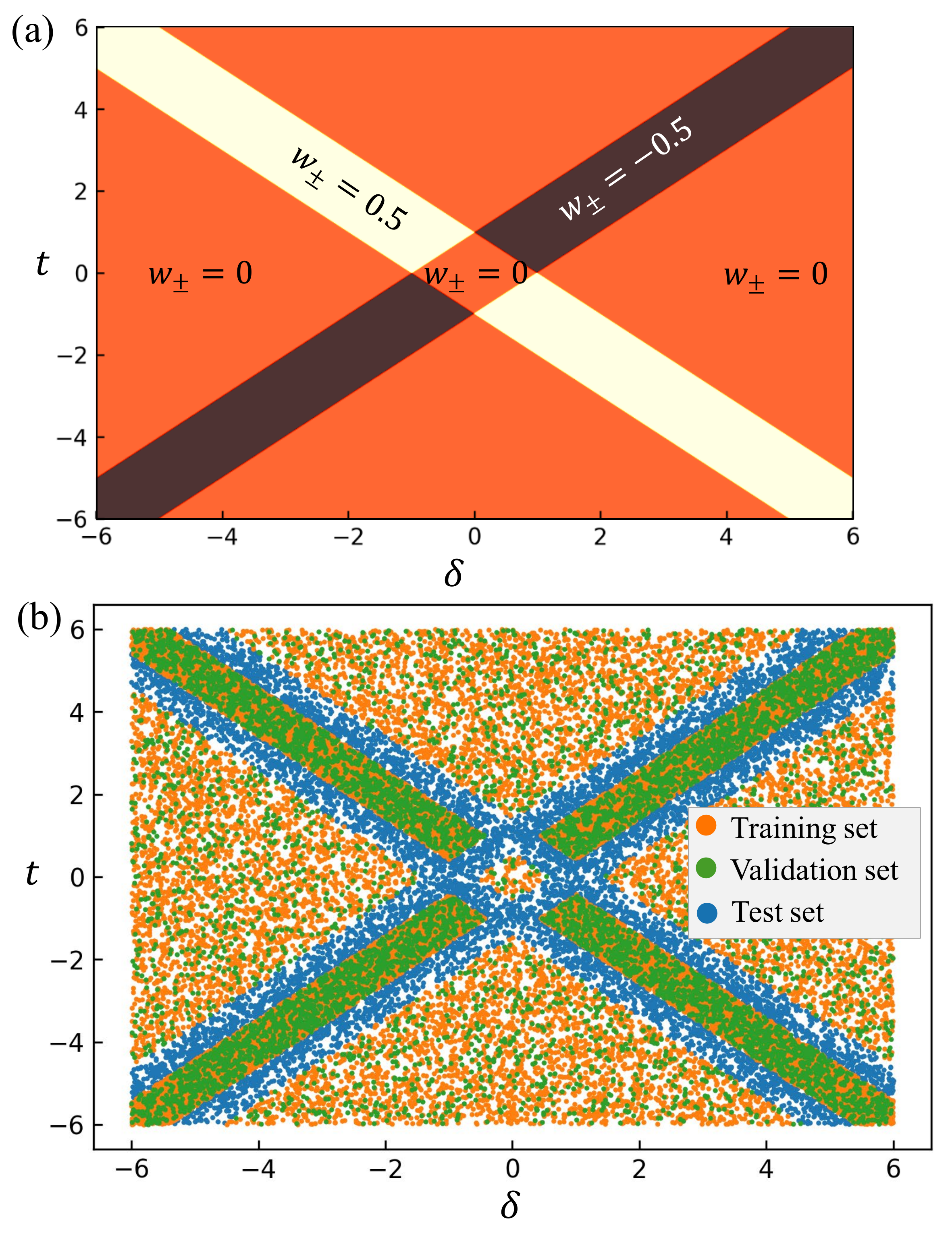}
    \caption{(Color online) (a) Phase diagram of the non-Hermitian SSH model for $t\in[-6,6]$, $\delta \in [-6,6]$, and $t'=1$. (b) Data set distribution when $T=0.2$; the amounts of the training data set, validation data set, and test data set are about $2.4\times10^4$, $6\times10^3$, and $6\times10^3$, respectively.}
    \label{fig8}
\end{figure}
\section{Training details} \label{app}

We first describe some training details for the Hatano-Nelson model. We use the deep learning framework PyTorch \cite{paszke2019pytorch} to construct and train the neural network. Weights are randomly initialized to a normal distribution with the Xavier algorithm \cite{glorot2010understanding} and the biases are initialized to 0. We use the Adam optimizer\cite{kingma2014adam} to minimize the output of the neural network $\tilde{w}$ with the true value $w$. We set the initial learning rate at 0.001 and use the ReduceLROnPlateau algorithm \cite{paszke2019pytorch} to lower by 10 times when the improvement of the validation loss stops for 20 epochs. All hyper-parameters are set to default, unless mentioned otherwise. In order to prevent neural overfitting, $L_2$ regularization with strength $10^{-4}$ and early stop \cite{yao2007early} are used during the training. We use mini-batch training with the batch size 64 and a validation set to confirm that there is no overfitting during training. We take $4\times 10^3$ configurations, which consist of $1:1:1$ samples with winding numbers $w=\pm \{1,2,3\}$. The typical loss during a training instance of the CNN and FCNN is shown in Fig. \ref{fig7} (a), from which one can see that there is no sign of overfitting.

We now provide some training details for the non-Hermitian SSH model. In this case, the CNN has two convolution layers with 32 kernels of size $1\times2 \times 2$ and 1 kernel of size $ 32\times1 \times 1$, followed by a fully connected layer of 16 neurons before the output layer. In this model, the output layer consists of three neurons for three different inter-band winding numbers. All the hidden layers have ReLU as activation functions and the output layer has the softmax function $ f(\mathbf{x})_i=\exp{\mathbf{x}_i}/\sum_{j=1}^{n}\exp{\mathbf{x}_j}$. The exact topological phase diagram in the parameter space spanned by $t$ and $\delta$ is shown in Fig. \ref{fig8} (a). The training data set satisfying $s\geq T$ with $T=0.2$ here and the test data set satisfying $s< T$ are randomly sampled from the parameter space. The data set distribution is shown in Fig.~\ref{fig8} (b). The numbers of configurations in the training data set, validation data set, and test data set are about $2.4\times10^4$, $6\times10^3$, and $6\times10^3$, respectively. Typical loss during training instances of the CNN for different training data sets is plotted in Fig. \ref{fig7} (b), which clearly shows that the neural networks converge quickly without overfitting.

Finally, we present briefly some details for the non-Hermitian generalized AAH model. In this case, the validation set consists of $8\times10^3$ configurations corresponding to non-reciprocal-hopping Hamiltonians (with $h=0$) that are not included in the training data set. The typical loss is shown in Fig. \ref{fig7} (c), with the networks converging quickly without overfitting.

\begin{acknowledgments}
We thank Dan-Bo Zhang for helpful discussions. This work was supported by the National Natural Science
Foundation of China (Grants No. U1830111, No. U1801661, and No. 12047522), the Key-Area Research and Development Program of Guangdong Province (Grant No. 2019B030330001), and the Science and Technology Program of Guangzhou (Grants No. 201804020055 and No. 2019050001).
\end{acknowledgments}

\bibliography{reference}

\end{document}